\documentclass[prx,twocolumn,aps,epsf,showpacs,superscriptaddress]{revtex4-1}
\usepackage[pdftex]{graphicx}

\usepackage{dcolumn}
\usepackage{bm}
\usepackage{epsfig}
\usepackage{latexsym} 
\usepackage{amsmath}
\usepackage{amssymb}
\usepackage{color}
\usepackage{array}
\usepackage{bbm}
\usepackage{hyperref}
\usepackage{color}
\usepackage{array}
\usepackage{cancel}
\usepackage{ulem}

\def\8{\infty}

\def\undertext#1{\vtop{\hbox{#1}\kern 1pt \hrule}}

\def\pd#1{\partial_{#1}}

\def\be{\begin{equation}}
\def\ee{\end{equation}}
\def\bea{\begin{eqnarray} & &}
\def\eea{\end{eqnarray}}

\newcommand{\e}{\varepsilon}

\renewcommand{\(}{\left(}
\renewcommand{\)}{\right)}
\renewcommand{\[}{\left[}
\renewcommand{\]}{\right]}
\renewcommand{\v}[1]{\boldsymbol{#1}} 
\renewcommand{\vec}[1]{\boldsymbol{#1}}

\renewcommand{\pd}{\partial}

\newcommand{\header}[1]{\vspace{4pt}\noindent{\bf #1 -- }}

\graphicspath{{./Figures/}}

\usepackage[pdftex]{graphicx}
\usepackage{bm}
\usepackage{hyperref}

\newcommand{\beq}{\begin{equation}}
\newcommand{\eeq}{\end{equation}}

\begin{document}

\title{Quantum Hall network models as Floquet topological insulators}

\author{Andrew C. Potter}
\affiliation{Department of Physics, University of Texas at Austin, Austin, TX 78712, USA}
\author{J. T. Chalker}
\affiliation{Theoretical Physics, University of Oxford, Parks Road, Oxford OX1 3PU, United Kingdom}
\author{Victor Gurarie}
\affiliation{Department of Physics and Center for Theory of Quantum Matter, University of Colorado, Boulder, Colorado 80309, USA}

\begin{abstract}
Network models for equilibrium integer quantum Hall (IQH) transitions are described by unitary scattering matrices, that can also be viewed as representing non-equilibrium Floquet systems. The resulting Floquet bands have zero Chern number, and are instead characterized by a chiral Floquet (CF) winding number. This begs the question: How can a model without Chern number describe IQH systems? We resolve this apparent paradox by showing that non-zero Chern number is recovered from the network model via the energy dependence of network model scattering parameters. This relationship shows that, despite their topologically distinct origins, IQH and CF topology-changing transitions share identical universal scaling properties.
\end{abstract}

\pacs{
}
\maketitle

Disorder and localization often play a vital role in stabilizing topological matter. In particular, these features are essential for the experimental observation of quantized Hall conductance in a variety of experimental systems such as GaAs quantum wells~\cite{Li:2005de,Li:2009de}, graphene~\cite{giesbers2009scaling}, and magnetically doped topological insulator thin films~\cite{chang2013experimental}. These systems all allow tuning between Hall plateaus with different quantized conductance via changing gate voltages or external magnetic fields, resulting in a topological phase transition. 
This transition is marked by a jump in the Chern number of the occupied electron states and a change in the number of chiral edge states circulating around the perimeter of the sample. It is accompanied by a divergence in the localization length at a critical value of the chemical potential. 

Disorder and localization can also stabilize driven systems against drive-induced heating, giving access to new regimes of quantum coherent dynamics. Recent theoretical advances have shown that time-periodically driven (Floquet) systems can exhibit new types of non-equilibrium topological phases with inherently dynamical properties that could not arise in the ground state of static (time-independent) Hamiltonians~\cite{kitagawa2010topological,Rudner:2013bg,roy2017periodic}. Striking examples include chiral Floquet (CF) phases~\cite{Rudner:2013bg}, which exhibit chiral edge states, despite having only topologically trivial bulk bands. 

Although CF edge states are reminiscent of those in IQH systems, there are important differences. Crucially, the number of CF edge states for a given Floquet operator is the same at all values of the (compact) quasi-energy. By contrast, the number of edge states for a (time-independent) IQH Hamiltonian is a function of (non-compact) energy, and is given by the net Chern number of all bulk states at lower energies.
As a corollary, CF phases differ from IQH phases in that they do not exhibit charge pumping through bulk states upon adiabatic insertion of magnetic flux, and hence have vanishing Hall conductance and Chern number. Instead, for non-interacting systems, the unitary time-evolution operators for CF phases are characterized by an integer-valued winding number, $\chi$, the chiral unitary invariant~\cite{Rudner:2013bg}.

In this paper, we explore a relation between these two distinct types of topological phenomena via the network model introduced by Chalker and Coddington~\cite{Chalker1988} to describe the scattering dynamics of electrons near the quantum Hall plateau transition. 
The Chalker-Coddington network model (CCN) is defined by a unitary matrix that acts on a wavefunction sampled at discrete spatial lattice points in a continuum IQH system. 
This unitary can also be interpreted as the Floquet operator 
of a time-periodically driven system~\cite{klesse1999modeling}, as previously discussed in the context of photonic networks where it was found that these network models could realize both CF phases and Chern bands for appropriate network geometries and parameters~\cite{pasek2014network,delplace2017phase,delplace2019topological}.
We  construct a periodically time-dependent Hamiltonian, acting on the same lattice, whose Floquet operator coincides with the CCN unitary, and demonstrate that this Floquet system hosts a CF phase, and find vanishing Chern number for all bands at all fixed network-model parameters.

This correspondence uncovers a puzzle: IQH systems are characterized by a Chern number but  the CF system has Chern number zero.
We resolve this puzzle by showing that the Chern number of an IQH system is recovered by properly accounting for energy-dependence of scattering phases and tunneling amplitudes in the model. This relationship provides a fresh perspective on IQH systems, building on the fact that the CF Floquet operator is characterized by the chiral unitary invariant. Since the CCN is represented by the same unitary matrix as a CF system, values of the chiral unitary invariant can be used to label IQH phases within the CCN description. 

These observations show that the network model equally describes the fixed-energy scattering behavior of a wave-packet in an IQH system, and the full dynamics of a CF system. In particular, this implies that the topology changing phase transitions for disordered, non-interacting IQH and CF phases share the same universal scaling properties, in agreement with recent field-theoretic analysis~\cite{woo2019quantum} relating sigma-models for CF and IQH phases.

\header{Network model definition}
Consider the ground state of a non-interacting electron gas in a uniform magnetic field and a disordered potential, with the lowest Landau level partially occupied on average. If the potential is smooth on the scale of the magnetic length and has fluctuations smaller in amplitude than the cyclotron energy, the system is divided into spatial regions around potential minima where the Landau level is locally fully occupied, and regions around maxima where it is empty. Chiral edge states at the chemical potential circulate along the boundaries between these occupied and empty regions. Tunneling between distinct edge segments occurs near saddle-points in the potential where their spatial separation is small.

\begin{figure}[tb]
\centering
\includegraphics[width=0.5\textwidth]{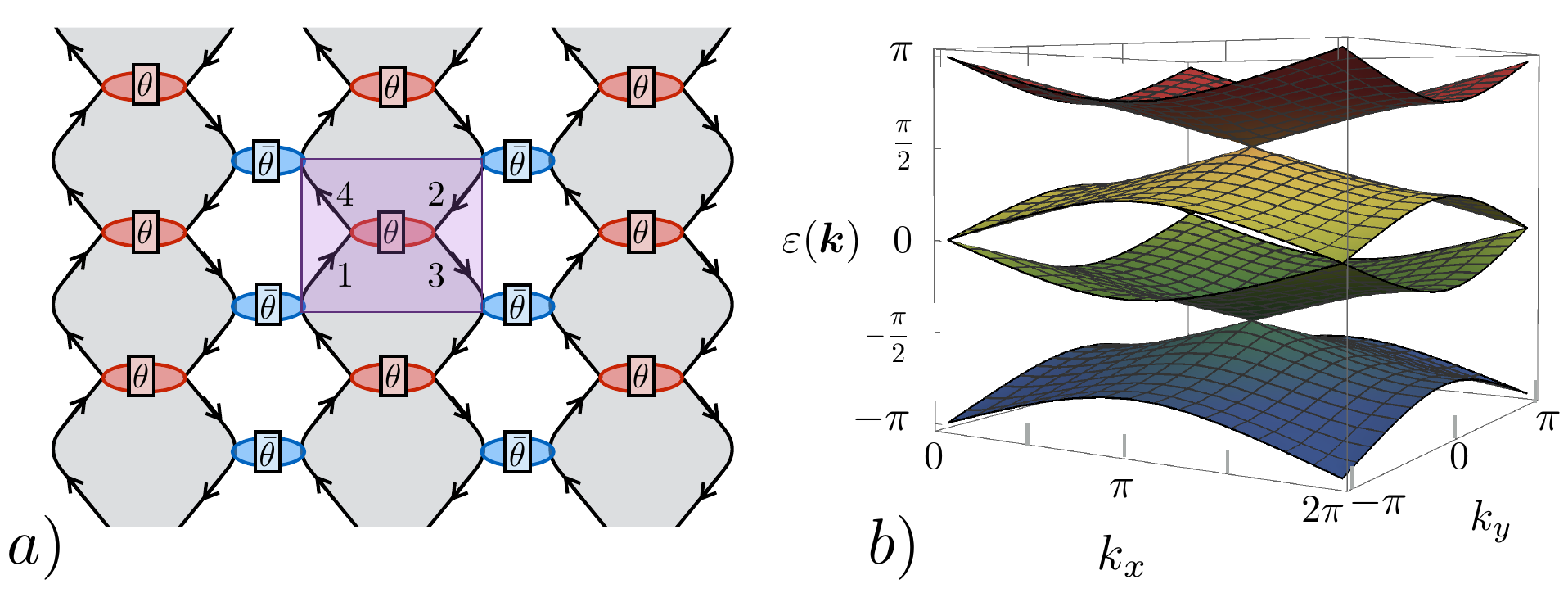}
\caption{{\bf Network model and Floquet band structure -- } a) schematic of Chalker-Coddington network model, b) Floquet bands $\e(\v{k})$ of the clean network model evolution operator at criticality, $\theta=\pi/4$.}
\label{fig:bands} 
\end{figure}

The network model~\cite{Chalker1988} describes a simplified version of this picture, in which the potential is chosen so that occupied and empty regions form alternate plaquettes of a regular square lattice. Edge states propagate on directed links of this lattice, meeting at nodes that correspond to potential saddle-points. Consider a stationary state in this continuum problem. 
Its wavefunction is sampled at a single point $r_i$ on each link, and represented by a current amplitude $\psi_i$. Amplitudes on incoming and outgoing links at a node are related by a scattering matrix, so that (referring to Fig.~\ref{fig:bands}):
\begin{align}\label{eq:node}
	\begin{pmatrix} 	\psi_3 \\ \psi_4 \end{pmatrix} 
	= \begin{pmatrix} e^{i\varphi_3} & 0\\ 0 & e^{i\varphi_4} \end{pmatrix} 
	\begin{pmatrix} 
		\cos \theta & \sin\theta 
		\\ -\sin\theta & \cos\theta 
	\end{pmatrix} 
	\begin{pmatrix}\psi_1 \\ \psi_2 \end{pmatrix}.
\end{align}
Here, $\theta$ parameterizes tunneling while $\varphi_3$ and $\varphi_4$ are Aharonov-Bohm phases.
A similar $S$-matrix with $\theta\rightarrow \bar\theta \equiv \pi/2-\theta$ describes scattering at the $\bar\theta$ nodes. Disorder is modeled by taking  $\varphi_i$ to be an independent random variable on each link $i$, uniform in $[0,2\pi)$. The model can be specified  for a closed system of $N$ links by an $N\times N$ unitary matrix $U$, which is composed of $2 \times 2$ sub-matrices, each having the form of Eq.~\ref{eq:node}~\cite{KlesseMetzler}. 

An eigenstate of the continuum Hamiltonian, sampled at the points $r_i$, is represented by an eigenvector of $U$ with eigenphase zero (mod $2\pi$), 
such that Eq.~\eqref{eq:node} is satisfied at every node. 
In order to find the energies at which scattering eigenphases vanish, and hence relate eigenvectors of $U$ to Hamiltonian eigenstates, it is necessary to consider the energy dependence of $\theta$ and $\varphi_i$.
From the shape of equipotentials near a saddle-point, one sees that $\theta$ increases from $0$ to $\pi/2$ as energy is increased across the disorder-broadened Landau level.
The accumulated link phase around a plaquette is ($2\pi$ times) the number of flux quanta passing through this plaquette. Randomness in $\phi_i$ arises from small random variations in the area of plaquettes, and energy-dependence of $\phi_i$ arises due to to the change in area of a plaquette with a change in the chemical potential.

The behavior of the model is simplest at the extreme limits $\theta=0$ and $\theta=\pi/2$. In the first case, the system consists only of isolated plaquettes enclosing occupied regions. In the second case, it is made up of isolated plaquettes enclosing empty regions, together with a chiral edge state at a boundary where the system meets an external empty region. Detailed numerical studies show that the edge state is present for all $\theta > \pi/4$.

\header{Floquet perspective}
These scattering dynamics define a \emph{stroboscopic Floquet evolution} in which we introduce a discrete time variable $t$ and take
\be 
\psi(t+1) = U \psi(t),
\ee
where $\psi(t)$ is the vector of all the link amplitudes at the time step $t$~\cite{Ho1996}. Writing the eigenvalues of $U$ as $e^{-i \e}$, the phases $\e$ are referred to in this context as quasi-energies, located within a compact Brillouin zone $\e \in (-\pi, \pi]$. The fact that quasi-energies lie on a circle rather than an open line changes the topological classification of Floquet dynamics compared to that of gapped ground-states of static Hamiltonians~\cite{kitagawa2010topological,Rudner:2013bg,roy2017periodic}. 

Non-interacting Floquet bands in systems with conserved particle number are exhaustively classified by two integer-valued topological invariants: 
\begin{enumerate}
\item the Chern number, $C_n$, defined separately for each Floquet band, $n$, and 
\item the chiral unitary index, $\chi(U)$, defined for the full Floquet unitary, which characterizes the number of chiral edge states that wrap around the quasi-energy Brillouin zone~\cite{Rudner:2013bg}.
\end{enumerate}

The Chern numbers for the Floquet-bands of $U$ are identically zero \cite{pasek2014network,delplace2017phase}. This can be verified by inspection at the trivial and topological limiting points, $\theta=0$ and $\theta=\pi/2$. At these points, the bulk motion consists of short loops around the elementary plaquettes of the network model (Fig.~\ref{fig:bands}), indicating that the Floquet bands admit a strictly localized Wannier basis of orbitals, each having support only on the four links of a single plaquette. Such a localized Wannier basis implies vanishing Chern number~\cite{thouless1984wannier}, immediately implying that each individual eigenstate has zero Chern number.  This conclusion extends to all values of $\theta\neq \pi/4$, since Chern number may only change at a delocalization transition, which in this model occurs only at $\theta=\pi/4$. The regions either side of this critical point inherit the vanishing Chern number of their limiting points, at $\theta=0,\pi/2$ respectively.

This argument, valid for arbitrary disorder, can be directly verified for the clean version of the model by explicit 
computation of the Floquet band-structure. Introducing crystalline momenta $\vec{k}$, the matrix $U$ is block-diagonal with blocks of the form
\be\label{eq:uk}
U(\theta,\vec{k}) = \left( \begin{matrix} 
0 & 0 & \sin\theta e^{- i k_x} & \cos\theta e^{-ik_y} \cr
0 & 0 & {-\cos\theta e^{ik_y}} & {\sin\theta} e^{i k_x} \cr
\cos\theta & \sin\theta  & 0 & 0 \cr
-\sin\theta & \cos\theta & 0 & 0
\end{matrix} \right).
\ee
We denote the eigenvalues of $U(\theta,\vec{k})$ by $\e_n(\theta,\vec{k})$ for $n\in \{0,1, 2, 3\}$. 
At $\theta=0$ and $\theta=\pi/2$, $\e_n$ are independent of $\vec{k}$, taking the values $ \e_n=\pi/4+{\pi n}/{2}$.  As $\theta$ deviates from these extreme values, the bands disperse, until they touch at $\theta=\pi/4$ in a sequence of Dirac points at wave-vectors $\vec{k} = (0,0)$ or $\vec{k}=(\pi,\pi)$ (Fig.~\ref{fig:bands}). Tuning away from $\theta=\pi/4$,  the Dirac points develop a mass-gap with an alternating sign mass for Dirac points separated in quasi-energy by $ \Delta \e = \pi/2$. Equivalently, viewed in the three-dimensional parameter space $(\theta,\vec{k})$, these degenerate points form monopole sources of Berry flux, with overall cancelling monopole charge, resulting in vanishing net Chern number for all $\theta$.

\header{Chiral Floquet invariant}
Viewing the network model as a lattice Floquet system, 
since $U$ has only topologically trivial bulk bands, any non-trivial topological behavior must emerge from a non-trivial chiral unitary invariant, $\chi\neq 0$. As a first step, we compare behavior at $\theta=0$ and $\theta=\pi/2$ for a system with open boundaries, where $\chi$ can be computed simply by counting the number of chiral edge states wrapping the quasi-energy Brillouin zone~\cite{Rudner:2013bg,pasek2014network}. There is no edge state in the first case, and one in the second case. The eigenvectors of $U$ corresponding to the edge state can be given explicitly. Let integer $j$ label links in order along the boundary, and for simplicity set all $\varphi_j=0$. Then  the vector with $\psi_j = e^{ikj}$ on edge links and $\psi_i =0$ on all other links is an (unnormalized) eigenvector of $U$ with quasi-energy $\varepsilon_\text{edge} = k$. This mode indeed wraps the quasi-energy Brillouin zone as $k$ varies,  and is manifestly absent in the other phase of the model ($\theta\approx0$), demonstrating that
\begin{align}
\chi(U_\text{CCN}(\theta))= 
	\begin{cases} 
		0 & 0\leq \theta < \pi/4 \\
		1 & \pi/4< \theta \leq \pi/2
	\end{cases}
\label{eq:chi}
\end{align}
and hence that the network model describes a Floquet chiral unitary index changing phase transition.

Although inspection of the stroboscopic edge motion is sufficient to give the value of bulk invariant $\chi$ via the bulk-boundary correspondence established in~\cite{Rudner:2013bg}, it is also reassuring to compute $\chi$ directly from the network model in the bulk by considering a system with periodic boundary conditions. Here, we face an obstacle: to compute $\chi$ from bulk behavior alone, it is not sufficient to examine stroboscopic times (for which the bulk motion is always trivial when all $C_n=0$). Instead, one must examine the micro-motion within a single period. The network model as formulated is blind to this micro-motion, and additional choices are needed to define it (though the final result will be independent of these choices).

Specifically we seek a continuous path in the space of unitary matrices, from the identity to $U$. This can be generated by
 a local Hamiltonian $H(t)$ that acts on the Hilbert space of the network model lattice for all times $0\leq t \leq T$ within the Floquet period $T=1$. That is, we require $U(t) = \mathcal{T}e^{-i\int_0^tH(t_1)dt_1}$ (where $\mathcal{T}$ denotes time ordering) such that $U(T) = U$, the CCN unitary. This relation does not uniquely fix $H(t)$.
However, if we demand that $H(t)$ is spatially local, any such choice of $H(t)$ will produce the same value of $\chi$. In Appendix.~\ref{app:ht}, we construct a specific example via a coupled-wire perspective, and explicitly verify Eq.~\ref{eq:chi} via evaluation of the bulk chiral Floquet winding-number~\cite{Rudner:2013bg}. The completeness of the Floquet classification for lattice models~\cite{Rudner:2013bg,po2016chiral,fidkowski2019interacting} also implies that any such local generating $H(t)$ must be explicitly time-dependent (in contrast to the static Hamiltonian for continuum Landau levels).

%
\begin{figure}
\centering
\includegraphics[width=0.5\textwidth]{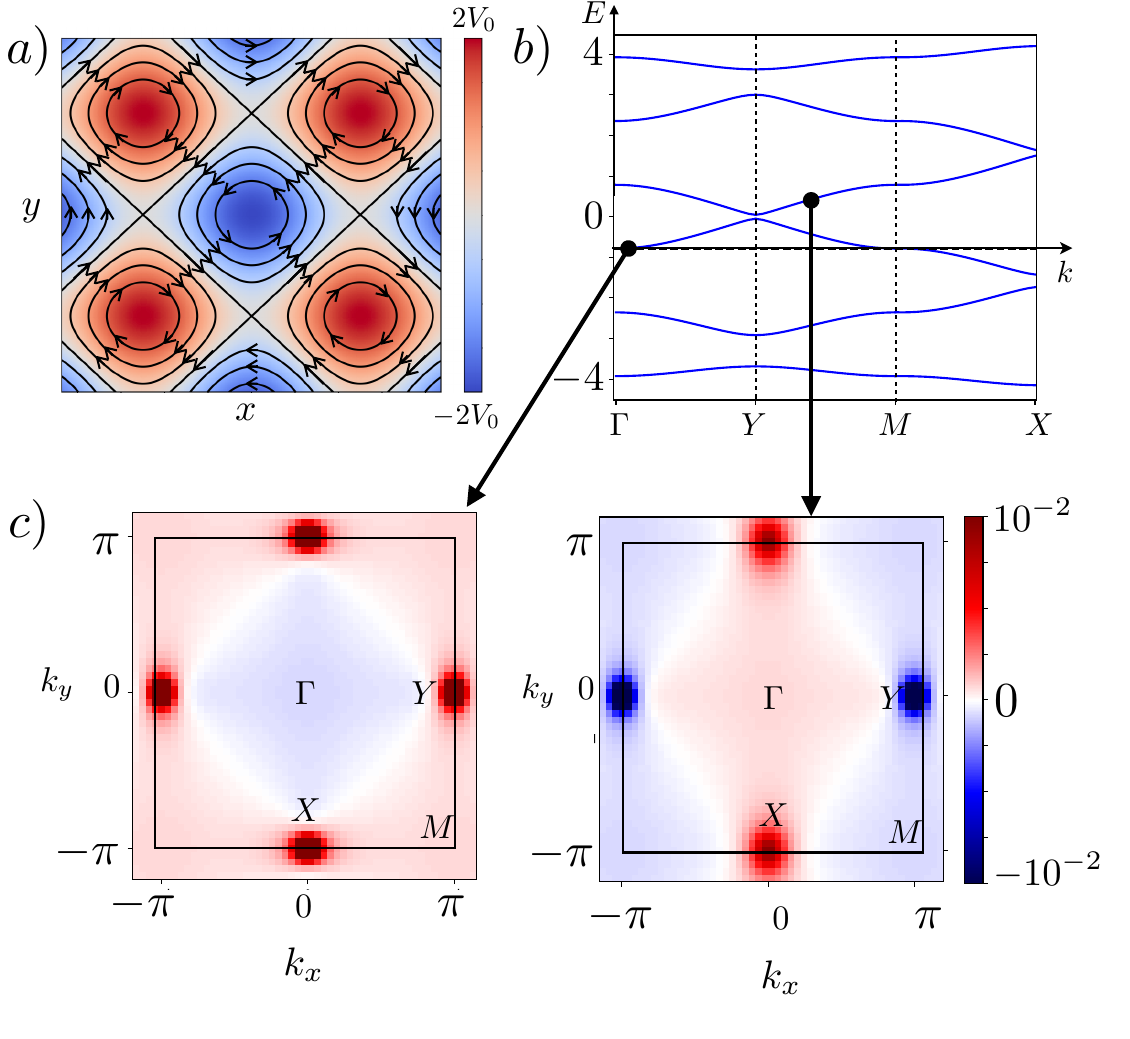}
\caption{{\bf Chern bands from the scattering network -- } a) Semiclassical orbits for an electron in a periodic potential and uniform magnetic field, b) Partial-set of energy bands from the scattering network with $\theta(E) = \frac{\pi}{4}\(\tanh\[\frac{E+\pi/4}{4\pi}\]+1\)$ and $\varphi(E) = E$. The horizontal axis is placed at the critical energy $E=E_c$ for which $\theta(E_c) = \pi/4$, c) Berry curvature for two bands near $E=0$ for which $\theta$ respectively crosses (left panel) and does not cross (right panel) $\pi/4$, giving total Chern number $1$ or $0$. All other bands have Chern number zero. 
}
\label{fig:bands2} 
\end{figure}

\header{Recovery of Chern bands}
These observations naturally raise the question: how can a model with zero Chern number describe the quantum Hall transition? We now show  that a non-zero Chern number for the Landau band of the continuum Hamiltonian is correctly recovered from the eigenvectors of $U$ when the network model parameters are allowed to vary with energy in a realistic manner. 

For clarity, we use the clean model in this discussion although the results are general. Without disorder, the link phases obey $\varphi_i = \varphi(E)$ for all $i$, where $\varphi(E)$ is a monotonic function of energy $E$ with an increment $\Phi$ across the energy width of the disorder-broadened Landau level. If there are many magnetic flux quanta per unit cell (the natural regime for the network model) then $\Phi \gg 1$. Including $\varphi(E)$, the quasi-energies of $U$ are $\e_n(\vec{k},\theta) - \varphi(E)$. Eigenenergies of the continuum problem form bands defined by:
\be\label{eq:self}
\varphi(E) = \e_n(\vec{k},\theta) + 2\pi m
\ee
for integer $m$. For large $\Phi$ there are many such bands.

Recalling that $\theta\equiv \theta(E)$ is a (slowly varying) function of $E$, the solution to Eq.~\eqref{eq:self} for each $m,n$ defines a surface $\theta_{m,n}(\vec{k})$ in the space $(\theta, \vec{k})$ that was introduced following Eq.~\ref{eq:uk}. For all but one of these surfaces, both of the oppositely charged $(\theta,\vec{k})$-space monopole sources of Berry flux lie on the same side of the surface. The net Berry flux through these surfaces is therefore zero, implying zero Chern number for the associated band of eigenstates in the continuum problem. However, there is one exceptional pair $(m,n)$ for which $\theta_{m,n}(\vec{k})<\pi/4$ at $\vec{k}=(0,0)$ but $\theta_{m,n}(\vec{k})>\pi/4$ at $\vec{k}=(\pi,\pi)$ (or vice-versa, depending on the parity of $n$). For this exceptional pair, oppositely charged monopoles lie on opposite sides of the surface, which therefore has a full unit of Berry flux passing though it. The associated band of eigenstates hence has unit Chern number, and summing over all bands we recover unit Chern number for the Landau level. A numerical demonstration for a particular energy dependence is shown in Fig.~\ref{fig:bands2}.

\header{Discussion} These two lines of analysis show that the network model equally describes both chiral Floquet topological insulators, and quantum Hall phases and transitions. Specifically, these results establish a precise equivalence of the dynamics of a wave-packet with near-constant energy in these two settings. Since the critical properties at topological phase transitions in both classes arise from delocalization at a fixed critical energy, this implies that these topologically distinct phenomena share the same critical properties at their (disordered) topological phase transitions. We emphasize that, despite sharing critical scaling properties the CF and IQH systems are sharply distinct. Beyond their distinct bulk-topology and edge-state structure, they exhibit qualitatively different dynamics of spatially localized wave-packets, which remain indefinitely localized in strongly-disordered CF phases~\cite{titum2016anomalous}, but instead spread sub-diffusively in IQH systems due to overlap with critically extended states~\cite{chalker1988scaling}.

 In the absence of interactions, the equilibrium classification and the Floquet classification have related structure~\cite{roy2017periodic}: for each equilibrium class with classification $G$, there is a corresponding set of purely dynamical Floquet phases also having classification $G$. Our results raise the question: are the critical properties of topological-phase transitions equivalent for all of these equilibrium/Floquet phase pairs? We conjecture that the answer is affirmative. For example, closely related $2d$ network models can be used to establish a similar relation between topological phase transitions in equilibrium and Floquet chiral superconductors with spin-rotation symmetry (spin-quantum Hall effect, class C) whose critical properties correspond to percolation~\cite{Kagalovsky,Gruzberg:1999jm,Senthil:1999fk}. Moreover, analogous $1d$ scattering network constructions can be obtained by compactifying $2d$ examples to extend these results to all $1d$ classes~\cite{toappear}, leading us to conjecture a general equivalence of universal scaling structure of non-interacting equilibrium and Floquet topological phase transitions.
 
The network model construction is special to non-interacting systems with elastic scattering. In interacting settings, the conjectured static/Floquet topological quantum criticality correspondence likely continues to hold for interacting MBL systems in $1d$ where phase transitions are characterized by RG-flow to infinite randomness and have equivalent interacting and non-interacting scaling properties~\cite{fisher1992random} By this route the $1d$ Floquet model studied in Ref.~\cite{berdanier2018floquet,BerdanierPRB} is related to an Ising model, which is known to have a network model representation \cite{Merz}. However, the correspondence will presumably fail for interacting $2d$ systems, since in that context CF phases are governed by rational-fractional-valued topological invariants with a completely distinct structure from Chern number~\cite{po2016chiral,fidkowski2019interacting}, and because the non-interacting CF-to-trivial critical-point will broaden into an intervening thermal phase upon including interactions~\cite{nandkishore2014marginal}.

\vspace{4pt}
\noindent{\it Acknowledgements -- } We thank Yang-Zhi Chou, Matthew Foster, Itamar Kimchi, Sid Parameswaran, and Romain Vasseur for insightful discussions. This work was supported by NSF DMR-1653007 (AP), by the EPSRC Grant No EP/S020527/1 (JTC), and by the Simons Collaboration on Ultra-Quantum Matter, which is a grant from the Simons Foundation (651440, VG).

\bibliography{CCN}

\appendix
\section{Time-dependent generating Hamiltonian for network models \label{app:ht}}
In this supplementary information, we build a local, time-dependent Hamiltonian that generates the network model dynamics.
In contrast to a previous construction relating Floquet and network model systems~\cite{delplace2019topological}
we use only the Hilbert space of the network model to produce a time-dependent Hamiltonian that directly generates the microscopic dynamics of an arbitrary network model, from which the bulk chiral Floquet invariant can be directly computed.

\begin{figure}[b!]
\includegraphics[width=0.5\textwidth]{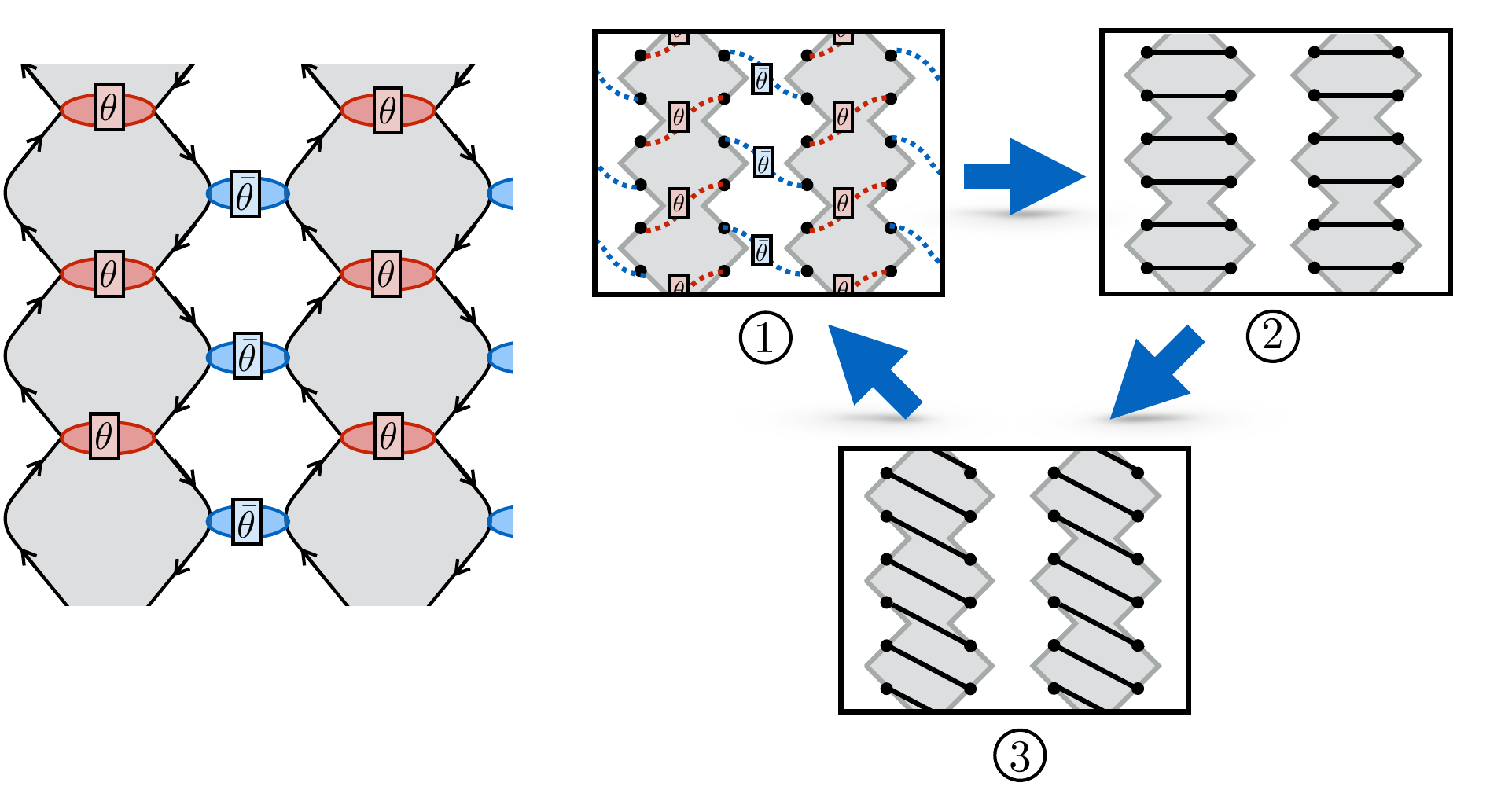}
\caption{{\bf Coupled wire picture of network model.}  Quasi-one dimensional strips with oppositely propagating edge states are indicated in gray. Red and blue scattering matrices with scattering angles $\theta$ and $\bar{\theta}=\pi/2-\theta$ represent inter- and intra- wire backscattering. Right panels 1-3: Each panel depicts the local couplings of a stroboscopic Hamiltonian that generates the network model dynamics. Solid black lines represent coupling $u_{ij}(\pi)$, dashed red and blue lines represent $u_{ij}(\theta)$ and $u_{ij}(\pi/2-\theta)$ respectively (see text for definitions).}
\label{fig:CCN} 
\end{figure}

To construct a generating $H(t)$, we adopt a coupled-wire perspective~\cite{Lee1994} by grouping the chiral segments of the model into pairs of alternating strips (Fig.~\ref{fig:CCN}). We build $H(t)$ from a sequence of piecewise time-independent Hamiltonians consisting of particle hoppings between disjoint pairs $\alpha,\beta$ of sites. These Hamiltonians are sums of terms of the form $h_{\alpha \beta} = J \sigma^y$, where $\sigma^y$ is the Pauli matrix acting on the two amplitudes $(\psi_\alpha, \psi_\beta)^{T}$.
When applied for time $t=\lambda/J$ this produces the unitary evolution $u_{\alpha \beta}(\lambda) = e^{-i \lambda \sigma^y}$.
Opposite chiral motion at the edge of each one-dimensional strip is produced in the second and third portions of the period by applying sequential SWAP operations, $u_{\alpha \beta}(\lambda=\pi)$, on the horizontal and diagonal bonds, shown on panels 2 and 3 of Fig.~\ref{fig:CCN}. 
Scattering is introduced via an initial partial swap step, by applying $u_{\alpha \beta}(\theta)$ and $u_{\alpha \beta}\(\frac{\pi}{2}-\theta\)$ at the inter- or intra- wire scattering nodes respectively (panel 1 of Fig.~\ref{fig:CCN}).

With a local generating Hamiltonian in hand, one can compute $\chi$ directly via the formula~\cite{Rudner:2013bg}
\begin{align}
\chi[\tilde{U}] &= \frac{1}{8\pi^2} \int dtdk_xdk_y \text{Tr}\(\tilde{U}^\dagger\pd_t\tilde{U} \[\tilde{U}^\dagger\pd_{k_x}\tilde{U},\tilde{U}^\dagger\pd_{k_y}\tilde{U}\]\)
\nonumber\\
\tilde{U}(t,\vec{k}) &= 
\begin{cases} 
U(2t,\vec{k})  & 0<t\leq T/2 \\
e^{+2iH_F(t-T/2)}U(T,\vec{k})  & T/2<t\leq T\,. 
\end{cases}
\label{eq:RLBL}
\end{align}
Here, $\tilde{U}(t)$ is the Floquet unitary time evolution $U(t,\vec{k})$, supplemented by backwards time-evolution under the time-independent Floquet Hamiltonian $U(T,\vec{k}) = e^{-iH_FT}$, to deform $U(t)$ into a unitary loop: $\tilde{U}(0)=\tilde{U}(T)=1$.
One can confirm:
\begin{align}
\chi(U_\text{CCN}(\theta))= 
	\begin{cases} 
		0 & 0\leq \theta < \pi/4 \\
		1 & \pi/4< \theta \leq \pi/2
	\end{cases}
\label{eq:chi}
\end{align} 
directly using Eq.~\ref{eq:RLBL}. 
This computation is simplest to do for $\theta=0,\pi/2$, where the network model unitary { raised to the fourth power} is already a unitary loop: $U^4(T, \vec k)=-1$. Appealing to the topological invariance under perturbations that do not close gaps at quasi-energies $0,\pi$, we can then extend this result to generic values of $\theta\neq \pi/4$.

A key requirement in this evaluation of $\chi$ is the construction of a closed, non-trivial loop in the space of unitaries, passing through the identity and the network model $U$. That construction is achieved here by the introduction of $H(t)$. An alternative approach to constructing such a loop, without direct reference to a Floquet Hamiltonian, is discussed in \cite{delplace2017phase}.

\end{document}